\documentclass[11pt,a4paper]{article}
\usepackage{amsmath,algorithm,bm,natbib,algorithm,caption,graphicx}
\usepackage[margin=1.0in]{geometry}

\title{Robust deblending of simultaneous source seismic data}

\author{Aaron Stanton and Keith Wilkinson \\ 
Key Seismic Solutions Ltd.\\
Calgary, Alberta}

\begin{document}

\maketitle

\begin{abstract}

Simultaneous source seismic acquisition is an efficient method of seismic surveying that can considerably reduce the cost of high density seismic acquisition. The method results in overlapping records, or interference, that must be removed prior to subsequent processing. Deblending methods typically rely on the incoherence of the blending noise relative to the underlying signal. There are many common situations where these assumptions break down, for instance, when the underlying signal contains noise or erratic amplitudes, or when shooting times are not sufficiently random. We present a robust inversion based deblending algorithm that can overcome these challenges.

\end{abstract}

\section*{Introduction}
Blended seismic acquisition reduces the time needed to acquire a seismic survey by allowing neighbouring shots to overlap in time and space. Numerous studies have shown the effectiveness of the approach to provide high trace density data at a reduced cost \citep{berkhout2008changing,howe2009independent,li2017aspects}. A number of approaches to process blended data have been proposed. \cite{dragoset20093d} apply conventional processing to blended data, relying on the power of stacking to attenuate much of the noise, while \cite{doi:10.1190/1.1892045} rely on the orthogonality of source encodings to separate data by correlation. Others have applied conventional noise attenuation methods to separate blended data \citep{gulunay2001seismic}. More recently, a number of inversion based methods have been proposed to deblend seismic data \citep{abma2015independent}. 

\begin{figure}
\centering
  \includegraphics[width=0.75\linewidth]{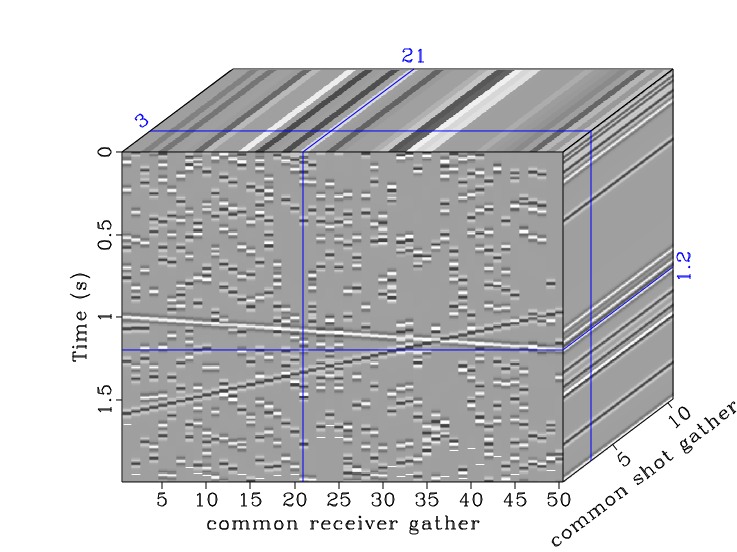}
  \caption{Illustration of the coherence property in deblending. Interference appears incoherent within a common receiver gather, and coherent within a common shot gather. }
  \label{fig:coherence}
\end{figure}

Inversion based deblending attempts to solve an underdetermined inverse problem. It is typically constrained by a regularization term that enforces coherency in the model. A variety of techniques have been used to enforce coherency in deblending, such as F-K filtering \citep{doulgeris2010separation}, median filtering \citep{van2012inversion}, rank reduction \citep{cheng2015separation}, as well as thresholding in the F-K and Radon domains \citep{abma2015independent, ibrahim2013simultaneous}. These methods all rely on the property that shot interference is incoherent in some domain (for example common receiver gathers). The incoherence of the interference depends on random relative shooting times between adjacent traces. If neighbouring traces in a domain are from the same source (for example in common shot gathers), the shot interference will be coherent from trace to trace and these methods will fail to isolate signal from interference. This property is illustrated in Figure \ref{fig:coherence}. The interference appears incoherent within a common receiver gather, and coherent within a common shot gather.

The underlying assumptions of inversion based deblending are as follows:
\begin{enumerate}
\item the unblended data is coherent
\item the interference is incoherent in some domain
\end{enumerate}
There are a number of reasons why assumption 1 could be violated, particularly in land acquisition. For example, variable source coupling, statics, surface waves, as well as survey noise all violate the assumption that the unblended data are coherent. It is important to make a distinction between survey noise and background noise. \cite{berkhout2013effect} point out that simultaneous shooting can provide a higher signal to noise ratio than conventional shooting simply because a shorter survey time will result in less recorded background noise relative to the total amount of signal imparted into the earth. While background noise may be reduced by this effect, in general we see an increase in the amount of survey noise in simultaneous shooting due to increased surface activity. Another obvious violation of assumption 1 are occasional shots with a vibrator malfunction. This could be a partial sweep or a shot that does not generate a start time. These shots represent a form of signal that clearly violates the blending system of equations. Lastly, spatio-temporal coherence in shooting patterns throughout a survey could unintentionally violate assumption 2. This can be particularly problematic for the deblending of low frequencies \citep{abma2014shot}. 

Violating these assumptions in the deblending process leads to loss of signal and poor attenuation of interference. We aim to overcome some of these challenges by incorporating robust statistics into the deblending problem.

\section*{Theory}

Blended seismic data, $\bm{d}$, can be modeled using unblended data, $\bm{m}$, via
\begin{equation}
\bm{d} = \bm{\Gamma} \bm{m},
\label{eq:fwd}
\end{equation}
where $\bm{\Gamma}$ is a blending operator that shifts and sums the data. Because $\bm{\Gamma}$ compresses the data, solving for $\bm{m}$ by minimizing $J = || \bm{d} - \bm{\Gamma} \bm{m} ||_2 $ is an ill-posed problem. The problem may be further constrained by including a regularization term that penalizes incoherent energy in the model, and a robust weighting function that mitigates the effects of outliers in the data
\begin{equation}
J = || \bm{W}(\bm{d} - \bm{\Gamma} \bm{m}) ||_2 + \mu|| \bm{D} \bm{m} ||_2,
\label{eq:cost_diff}
\end{equation}
where $\bm{W}$ is a diagonal matrix of weights and $\bm{D}$ is an operator that emphasizes sharp contrasts in the model (for example a spatial derivative matrix). While this objective function better constrains the problem while mitigating the effects of outliers in the data, it leads to two practical challenges. First, the operator $\bm{D}$ must be strong enough to penalize interference in the model, but gentle enough to recover subtle features in the model; and second, the operator $\bm{W}$ must be designed to suppress outliers in the data, but without the aid of spatial information ($\bm{d}$ represents blended data in continuous receiver gather format). We propose an alternative objective function,
\begin{equation}
J = || \bm{d} - \bm{\Gamma} \bm{m} ||_2 \; \; \text{subject to} \; \; \bm{m} = P_C\{P_E\{\bm{m} \} \},
\label{eq:cost}
\end{equation}
where $P_E\{\}$ is a projection that mitigates the effects of erratic amplitudes, and $P_C\{\}$ is a projection that enforces lateral coherency. This cascaded projection is able to enforce lateral coherency in the data while also overcoming the effects of erratic amplitudes and spatial aliasing that may result from coherency in shooting times. A number of different approaches can be used for the projection $P_E\{\}$ (for example median filtering), as well as for the projection $P_C\{\}$ (for example prediction filtering, rank reduction or thresholding in the Fourier, Radon, or Curvelet domains). To minimize Equation \ref{eq:cost} we use an accelerated gradient descent approach \citep{nesterov1983method}. 

\section*{Synthetic data example}
\begin{figure}
\raggedright
  \includegraphics[width=\linewidth]{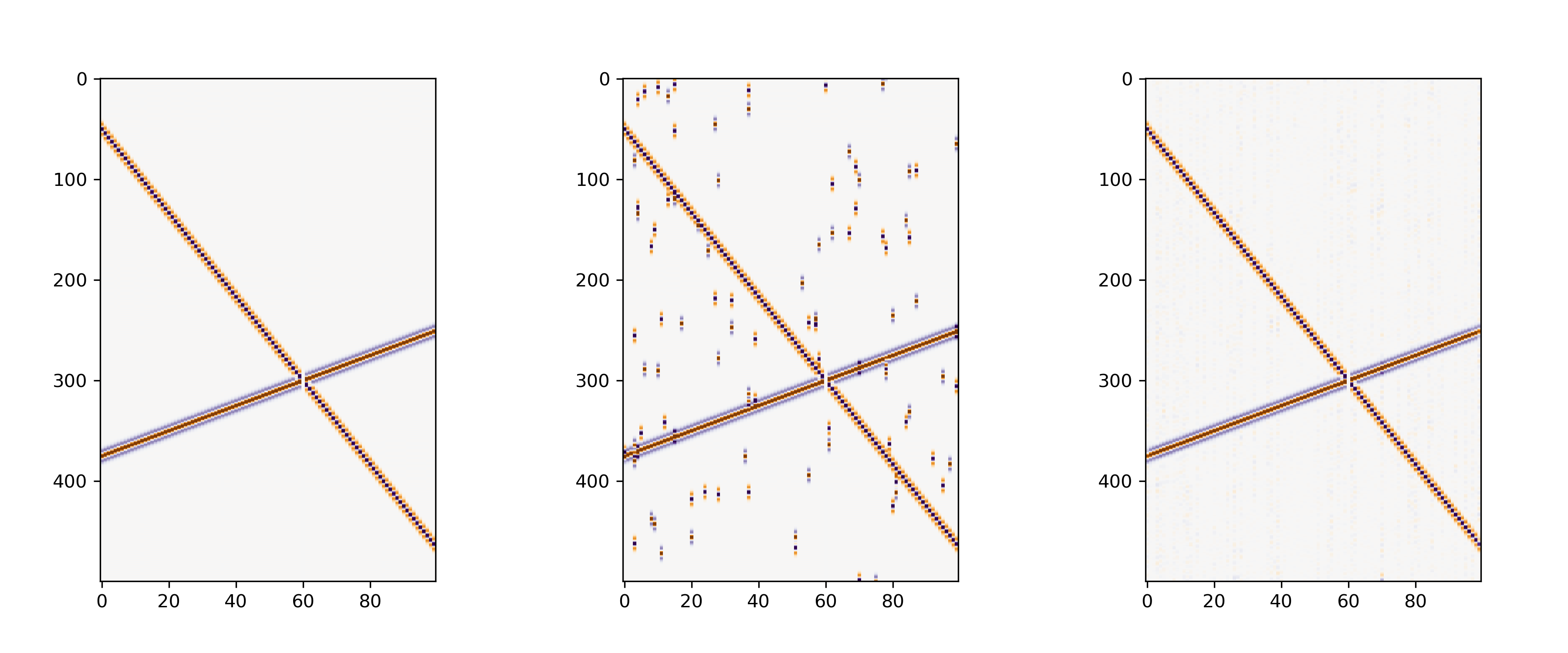}
  \caption{Deblending of noise free common receiver gather: true data (left), blended data (middle), and the result of deblending (right).}
  \label{fig:synth1}
\end{figure}

\begin{figure}
\raggedright
  \includegraphics[width=\linewidth]{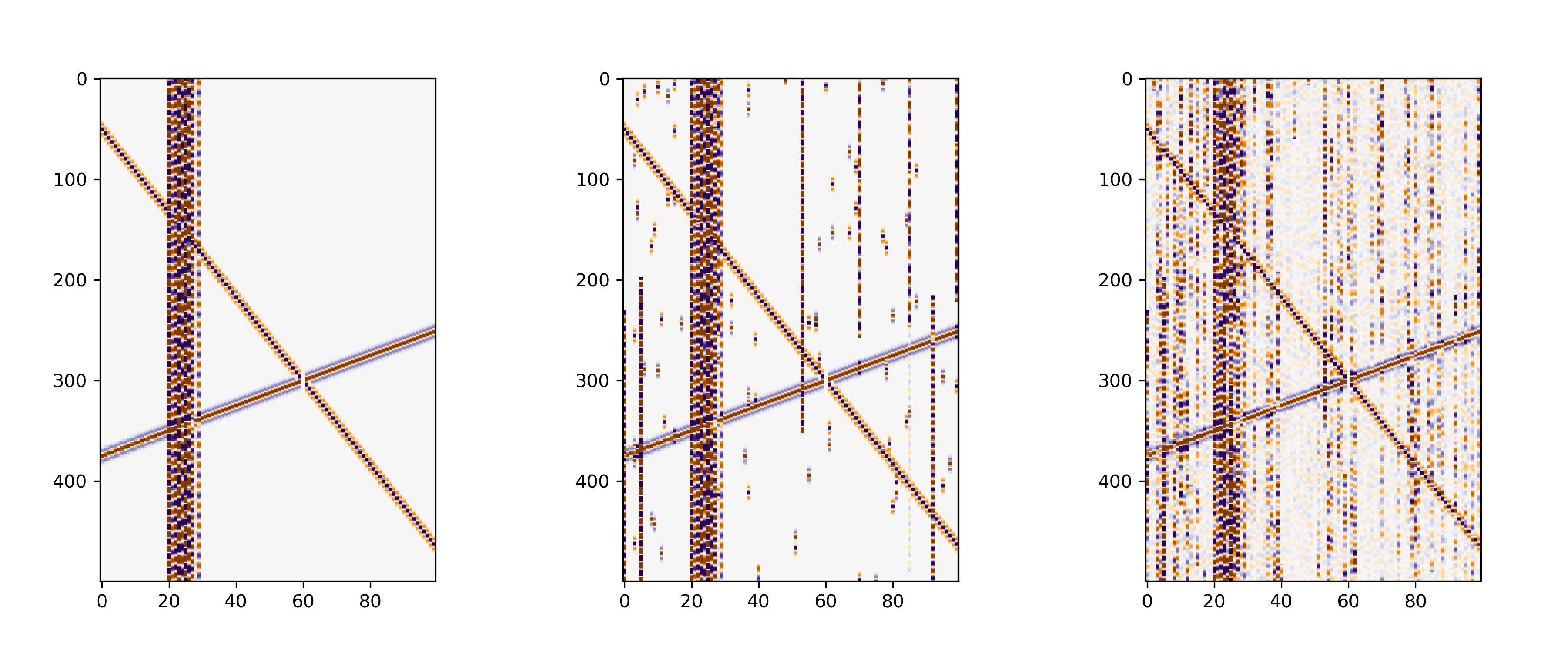}
  \caption{Conventional deblending of noise contaminated common receiver gather: true data (left), blended data (middle), and the result of conventional deblending (right).}
  \label{fig:synth2}
\end{figure}

\begin{figure}
\raggedright
  \includegraphics[width=\linewidth]{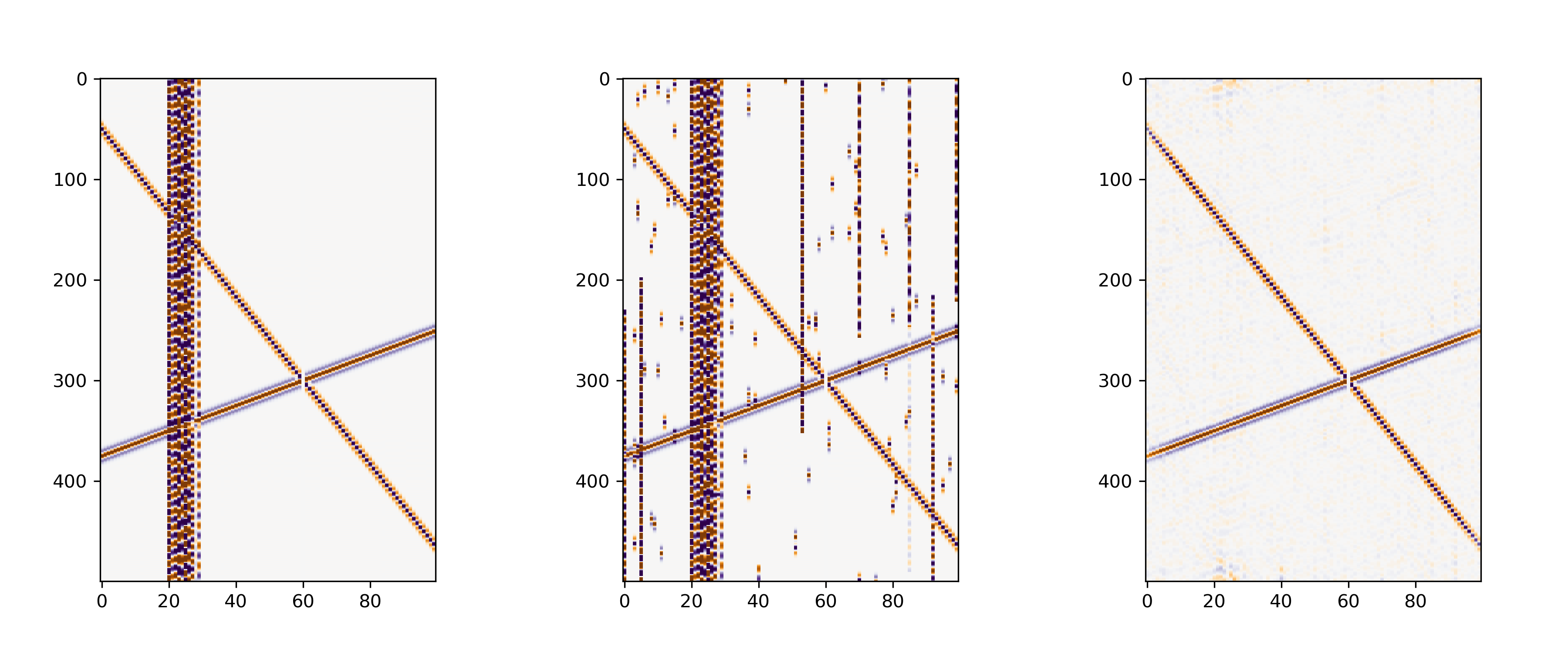}
  \caption{Robust deblending of noise contaminated common receiver gather: true data (left), blended data (middle), and the result of robust deblending (right).}
  \label{fig:synth3}
\end{figure}

To illustrate the effect of erratic amplitudes consider the synthetic data shown in Figure \ref{fig:synth1}. The data consist of a 100 trace common receiver gather with two dipping events (left panel). After blending the interference appears incoherent due to the delays between shots (middle panel). Deblending of these data using a conventional approach leads to satisfactory results (right panel). Now consider data with added erratic noise as shown in Figure \ref{fig:synth2} (left panel). After blending the data consist of blended signal as well as blended erratic noise (middle panel). After deblending (right panel) we see much of the interference resulting from coherent events has been attenuated, while the interference resulting from the erratic noise has been smeared throughout the data. Robust deblending (Figure \ref{fig:synth3}) is able to mitigate the effects of the outliers.

\section*{Field data example}
\begin{figure}
\centering
  \includegraphics[width=0.8\linewidth]{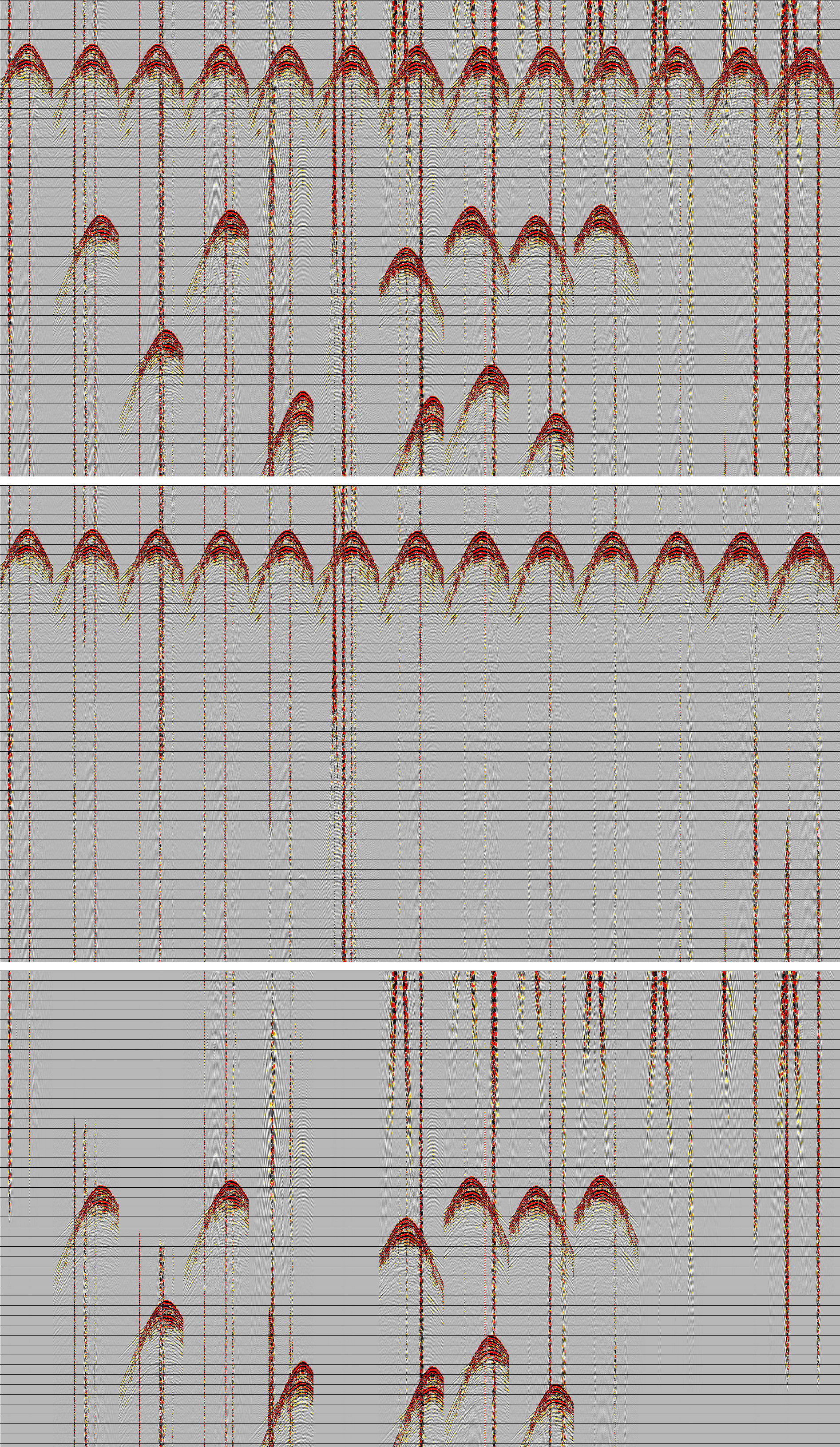}
  \caption{Shot gathers before deblending (top), after deblending (middle), and the difference (bottom).}
  \label{fig:field1}
\end{figure}

We applied deblending to a land dataset in the Western Canadian Sedimentary Basin. The data have an average blend fold of approximately 2 when measured over a +/- 5 second window. Figure \ref{fig:field1} (top) shows a selection of input shot records prior to deblending. The input data show several distinct types of blending interference including interference from earlier shots, interference from later shots, harmonic noise from later shots, as well as erratic noise. The data after deblending (middle) show a high level of blend noise attenuation, while the difference panel (bottom) shows a high level of signal preservation.

\section*{Conclusions}
Inversion based deblending is an effective method to supress simultaneous source interference, but special care must be taken to mitigate the effects of erratic amplitudes. In this abstract we presented a robust deblending approach that is able to achieve a high level of blend noise attenuation while preserving much of the underlying signal.


\section*{Acknowledgements}

We thank our colleagues Jeff Deere, Richard Bale and Mark Jiang for their advice and technical support and Key Seismic Solutions Ltd. for permission to publish. The seismic data used in the example was generously provided by an anonymous client.

\bibliographystyle{seg}  
\bibliography{paper}

\begin{thebibliography}{}
\itemsep0pt

\bibitem[Abma, 2014]{abma2014shot}
Abma, R.,  2014, Shot scheduling in simultaneous shooting, {\it in} SEG
  Technical Program Expanded Abstracts 2014: Society of Exploration
  Geophysicists,  94--98.

\bibitem[Abma et~al., 2015]{abma2015independent}
Abma, R., D. Howe, M. Foster, I. Ahmed, M. Tanis, Q. Zhang, A. Arogunmati, and
  G. Alexander,  2015, Independent simultaneous source acquisition and
  processing: Geophysics, {\bf 80}, WD37--WD44.

\bibitem[Berkhout, 2008]{berkhout2008changing}
Berkhout, G.,  2008, Changing the mindset in seismic data acquisition: The
  Leading Edge, {\bf 27}, 924--938.

\bibitem[Berkhout and Blacquiere, 2013]{berkhout2013effect}
Berkhout, G., and G. Blacquiere,  2013, Effect of noise in blending and
  deblending: Geophysics, {\bf 78}, A35--A38.

\bibitem[Cheng and Sacchi, 2015]{cheng2015separation}
Cheng, J., and M.~D. Sacchi,  2015, Separation and reconstruction of
  simultaneous source data via iterative rank reduction: Geophysics, {\bf 80},
  V57--V66.

\bibitem[Doulgeris et~al., 2010]{doulgeris2010separation}
Doulgeris, P., A. Mahdad, and G. Blacquiere,  2010, Separation of blended data
  by iterative estimation and subtraction of interference noise, {\it in} SEG
  Technical Program Expanded Abstracts 2010: Society of Exploration
  Geophysicists,  3514--3518.

\bibitem[Dragoset et~al., 2009]{dragoset20093d}
Dragoset, W., H. Li, L. Cooper, D. Eke, J. Kapoor, I. Moore, and C. Beasley,
  2009, A 3d wide-azimuth field test with simultaneous marine sources:
  Presented at the 71st EAGE Conference and Exhibition incorporating SPE
  EUROPEC 2009.

\bibitem[G{\"u}l{\"u}nay and Pattberg, 2001]{gulunay2001seismic}
G{\"u}l{\"u}nay, N., and D. Pattberg,  2001, Seismic interference noise
  removal, {\it in} SEG Technical Program Expanded Abstracts 2001: Society of
  Exploration Geophysicists,  1989--1992.

\bibitem[Howe et~al., 2009]{howe2009independent}
Howe, D., M. Foster, T. Allen, I. Jack, D. Buddery, A. Choi, R. Abma, T.
  Manning, and M. Pfister,  2009, Independent simultaneous sweeping in
  libya-full scale implementation and new developments, {\it in} SEG Technical
  Program Expanded Abstracts 2009: Society of Exploration Geophysicists,
  109--111.

\bibitem[Ibrahim and Sacchi, 2013]{ibrahim2013simultaneous}
Ibrahim, A., and M.~D. Sacchi,  2013, Simultaneous source separation using a
  robust radon transform: Geophysics, {\bf 79}, V1--V11.

\bibitem[Li et~al., 2017]{li2017aspects}
Li, C., C. Mosher, F. Janiszewski, Y. Ji, and S. Shaw,  2017, Aspects of
  implementing marine-blended source acquisition in the field, {\it in} SEG
  Technical Program Expanded Abstracts 2017: Society of Exploration
  Geophysicists,  42--46.

\bibitem[Martinez and Crews, 2005]{doi:10.1190/1.1892045}
Martinez, D.~R., and G.~A. Crews,  2005, Evaluation of simultaneous vibroseis
  recording, {\it in} SEG Technical Program Expanded Abstracts 1987: Society of
  Exploration Geophysicists,  577--580.

\bibitem[Nesterov, 1983]{nesterov1983method}
Nesterov, Y.~E.,  1983, A method for solving the convex programming problem
  with convergence rate {O} (1/k\^{} 2): Dokl. Akad. Nauk SSSR, 543--547.

\bibitem[van Borselen et~al., 2012]{van2012inversion}
van Borselen, R., R. Baardman, T. Martin, B. Goswami, and E. Fromyr,  2012, An
  inversion approach to separating sources in marine simultaneous shooting
  acquisition--application to a gulf of mexico data set: Geophysical
  Prospecting, {\bf 60}, 640--647.

\end{thebibliography}

\end{document}